\documentclass{jpp}
\usepackage{graphicx}
\usepackage{siunitx}
\usepackage{textcomp,gensymb} 
\usepackage{booktabs}
\usepackage{dcolumn}
\usepackage{bm}
\usepackage{makecell}
\usepackage[hyperindex=true,colorlinks=true,linkcolor=blue,citecolor=blue,urlcolor=blue]{hyperref}
\usepackage{amsmath}
\usepackage{amssymb}
\usepackage[utf8]{inputenc}
\usepackage[T1]{fontenc}

\shorttitle{particle acceleration in interpenetrating super-critical shocks}
\shortauthor{W. Yao, et al.}

\title{Investigating particle acceleration dynamics in interpenetrating magnetized collisionless super-critical shocks}

\author{
W. Yao\aff{1,2}
\corresp{\email{yao.weipeng@polytechnique.edu}},
A. Fazzini\aff{1},
S. N. Chen\aff{3},
K. Burdonov\aff{1,2,4},
J. B\'eard\aff{5},
M. Borghesi\aff{6},
A. Ciardi\aff{2},
M. Miceli\aff{7,8},
S. Orlando\aff{8},
X. Ribeyre\aff{9},
\and 
E. d'Humi\`eres\aff{9},
J. Fuchs\aff{1}
}

\affiliation{
\aff{1}LULI - CNRS, CEA, UPMC Univ Paris 06 : Sorbonne Universit\'e, Ecole Polytechnique, Institut Polytechnique de Paris - F-91128 Palaiseau cedex, France
\aff{2}Sorbonne Universit\'e, Observatoire de Paris, Universit\'e PSL, CNRS, LERMA, F-75005, Paris, France
\aff{3}``Horia Hulubei'' National Institute for Physics and Nuclear Engineering, RO-077125 Bucharest-Magurele, Romania
\aff{4}JIHT, Russian Academy of Sciences, 125412, Moscow, Russia
\aff{5}LNCMI-T, CNRS, Toulouse, France
\aff{6}Centre for Plasma Physics, The Queen’s University of Belfast, University Road BT71NN, Belfast, United Kingdom
\aff{7}Università degli Studi di Palermo, Dipartimento di Fisica e Chimica E. Segrè, 90134, Palermo, Italy
\aff{8}INAF, Osservatorio Astronomico di Palermo, 90134, Palermo, Italy
\aff{9}University of Bordeaux, Centre Lasers Intenses et Applications, CNRS, CEA, UMR 5107, F-33405 Talence, France
}

\begin{document}

\maketitle

\begin{abstract}
Colliding collisionless shocks appear in a great variety of astrophysical phenomena and are thought to be possible sources of particle acceleration in the Universe. We have previously investigated particle acceleration induced by single super-critical shocks (whose magnetosonic Mach number is higher than the critical value of 2.7) \citep{yao2021laboratory,yao2022detailed}, as well as the collision of two sub-critical shocks \citep{fazzini2022particle}. Here, we propose to make measurements of accelerated particles from interpenetrating super-critical shocks to observe the “phase-locking effect” \citep{fazzini2022particle} from such an event. This effect is predicted to significantly boost the energy spectrum of the energized ions compared to a single supercritical collisionless shock. We thus anticipate that the results obtained in the proposed experiment could have a significant impact on our understanding of one type of primary source (acceleration of thermal ions as opposed to secondary acceleration mechanisms of already energetic ions) of ion energization of particles in the Universe. 
\end{abstract}

\section{Introduction}

The particular origins of the high-energy particles (cosmic rays) flying through the Universe is still an open question. We know where they can possibly originate (pulsars, shocks, solar flares, magnetic clouds, etc.), but the nature of the acceleration mechanism(s) in each source, and hence the predictability of its occurrence, remains unconfirmed. 
Single shock structures are not the only place where particle acceleration is thought to take place. In fact, large-scale shock-shock collisions, with their different hydrodynamic and kinetic structures, are also seen as potential sources of high-energy particles.

Super-critical shock collisions can be found in OB associations (very young stellar objects) \citep{higdon2005ob,ackermann2011cocoon,rando2015fermi}, consisting of many hot giant stars and characterized by supernova (SN) explosions that are correlated in space and time \citep{higdon1998cosmic}. In these objects, the strong winds of giant stars and of SNs contribute to generating large-scale structures that are known as ``superbubbles''. Inside these structures are vast amounts of mechanical, thermal, and turbulent energy making them appealing sources of cosmic rays. Among the many structures inside these superbubbles include supercritical shock-shock collisions: wind-wind, SN-wind, and SN-SN shock collisions. The most common condition is the wind-wind shock collision in massive star binaries \citep{usov1992stellar,sanchez2019vlbi}, which is certainly a topic of large interest in astrophysics.

\section{Former Experimental Results}

\subsection{Single magnetized shock characterization}

In our former experiments performed at JLF/Titan and LULI2000 \citep{yao2021laboratory,yao2022detailed,fazzini2022particle}, we investigated single shock formation, as well as shock-shock interactions, by having an expanding plasma drive a shock into an ambient gas in the presence of a strong (20 T) external magnetic field. Since the expanding plasma, the ambient gas, and the magnetic field are all decoupled, we are able to vary the magnetic field independently of the expanding plasma and of the ambient gas, which allows us to investigate the essential role played by the magnetic field and to highlight the ion acceleration mechanism at play.

\begin{figure}
    \centering
    \includegraphics[width=\textwidth]{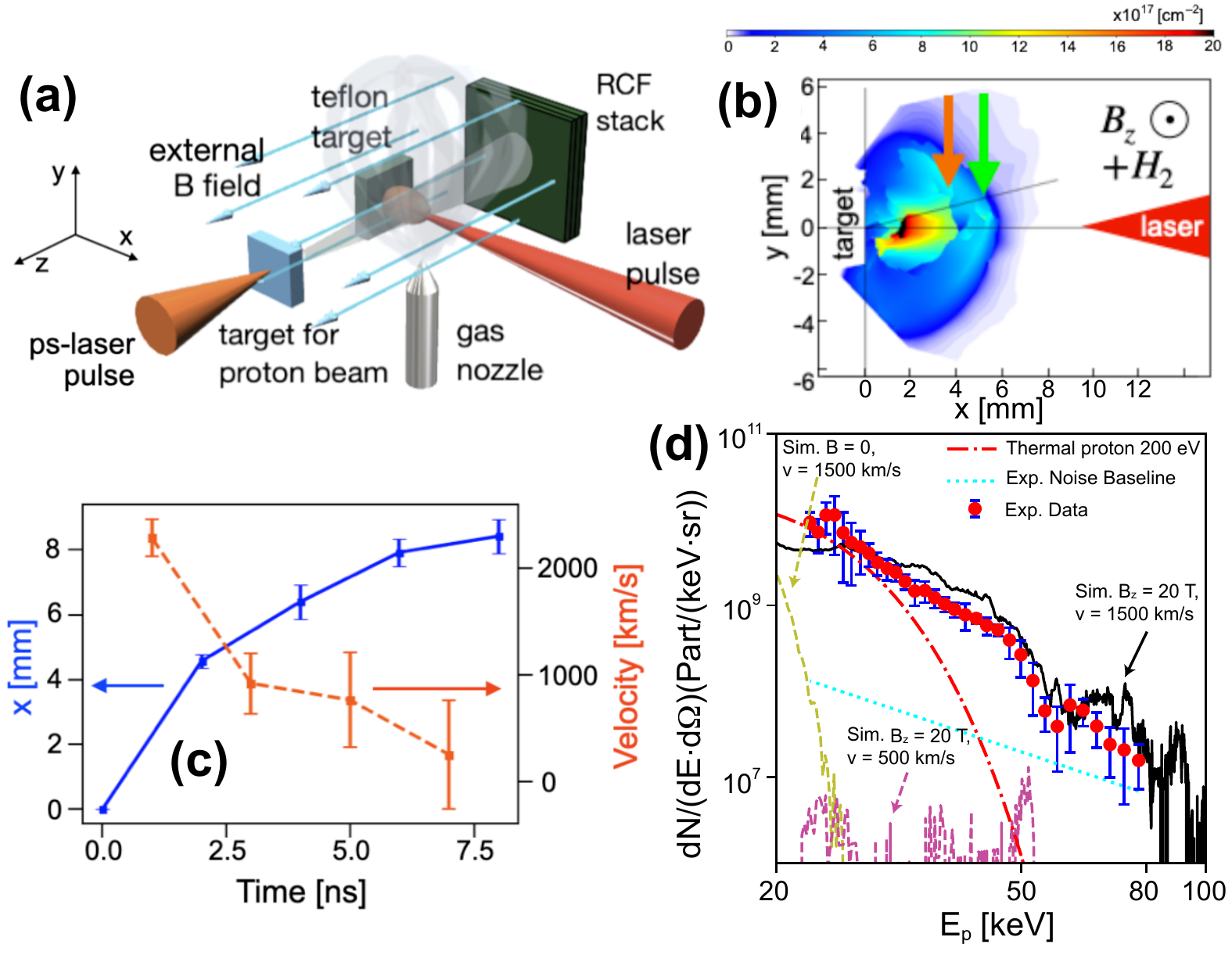}
    \caption{ (a) Setup of the experiment to characterize a single magnetized shock \citep{yao2021laboratory,yao2022detailed}. The whole scene is embedded in an H$_2$ gas of low density ($\sim 10^{18}$ cm$^{-3}$) emanating from a pulsed nozzle. Further, the whole assembly is embedded in a strong magnetic field (20 T). (b) Density measurement (integrated along the line of sight) 4 ns after the laser irradiation of the solid target. (c) Evolution of the shock front position along the x-axis and the corresponding velocity. (d) Evidence for the energization of protons picked up from the ambient medium. Proton energy spectra of both the experiment (red dots) and of three PIC simulations, the black solid line for the magnetized fast case with $B_z=20$ T and initial shock velocity $v=1500$ km/s, the yellow dashed line for the unmagnetized fast one with $B=0$ and $v=1500$ km/s, and the purple dashed line for the magnetized slow one with $B_z=20$ T and $v=500$ km/s, all measured at t=2.6 ns in the simulations. The analytical thermal proton spectrum is shown with the red dash-dotted line (200 eV); and the experimental noise baseline is shown in cyan dotted line. }
    \label{fig:setup}
\end{figure}

With the setup shown below in Fig.~\ref{fig:setup} (a), we have been able to characterize the plasma density, temperature, as well as the electric field developed at the front of a single shock, and observe strong non-thermal accelerated ion populations linked with the ``shock-surfing'' acceleration (SSA) mechanism \citep{yao2021laboratory,yao2022detailed}.

Specifically, the shock was generated by sending a plasma, generated from a high-power laser (1 ns duration, 70 J, $1.6\times 10^{13}$ W/cm$^2$) irradiating a solid target (Teflon, or CF$_2$), into a low-density gas (of density $\sim 10^{18}$ cm$^{-3}$), and in the presence of a 20 T magnetic field that is homogeneous and steady-state over the time scale of the experiment. This magnetic field is produced by our unique coil setup \citep{albertazzi2013production,higginson2019laboratory}. A record of the plasma electron density, obtained by optical probing, is shown in Fig.~\ref{fig:setup} (b). We can clearly observe the presence of an electron foot, likely reflected ahead of the shock by the moving shock front. To our knowledge, this is the first laboratory observation of such a structure, which was postulated \citep{dimmock2019direct} and also verified to be present in magnetospheric shock as recorded by satellite \citep{balogh2013physics}.

Figure~\ref{fig:setup} (c) shows the evolution of the shock front position and the corresponding velocity deduced from it, which shows the very fast decrease in shock velocity over the first few nanoseconds. Before 2.6 ns, the shock front velocity is around $v = 1500$ km/s, i.e. the shock is collisionless and super-critical with a mean-free-path larger than 10 mm and the magnetosonic Mach number larger than the critical value (see details in Table 1), while after around 5 ns, it becomes sub-critical, with correspondingly lesser energization of the reflecting ions from the ambient \citep{yao2021laboratory,yao2022detailed}. The ion spectrometer that was deployed in the experiment was set in the axis of the magnetic field, i.e., along the z-axis in Fig.~\ref{fig:setup} (a), otherwise the ions energized out of the plasma could not be recorded, as they would be deflected away by the 20 T large-scale magnetic field. 

The recorded proton spectrum is shown in Fig.~\ref{fig:setup} (d) by red dots evidenced non-thermal proton energization. We have performed a series of particle-in-cell (PIC) simulations, using the code SMILEI \citep{derouillat2018smilei}, with different magnetic field and initial shock velocity, which help us identify the critical role played by the external magnetic field and fast shock velocity in the non-thermal proton energization, i.e., only in the case with both 20 T magnetic field and 1500 km/s initial shock velocity can we well reproduce the experimental proton spectrum, otherwise the spectra will be lower than the noise baseline. Also note that there is no signal recorded above the noise baseline for cases with only the magnetic field or the ambient gas (i.e., when there will be no shock), indicating that the non-thermal particle populations are indeed coming from the shock.

\begin{figure}
    \centering
    \includegraphics[width=\textwidth]{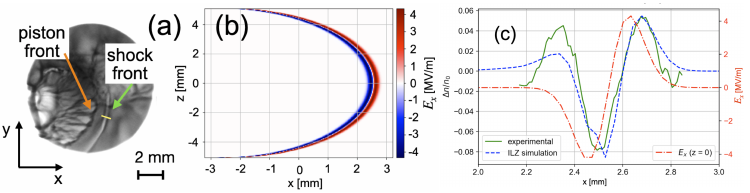}
    \caption{(a) Proton radiography (collected on a radiochromic film (RCF) and employing 19 MeV protons) of the same configuration as shown in Fig.~\ref{fig:setup} (a), 5 ns after the laser pulse. (b) Hemispherical electric field $E_x$ in the xz-plane, with a radius of $R=5.1$ mm, estimated from (a). (c) Lineout of the proton dose modulation along the yellow line indicated in (a). The green full curve is the modulation from the experimental results, and the dashed blue curve is that from the ILZ simulation \citep{bolanos2019laboratory}, which is obtained by imposing a bipolar electric field with hemispherical shape shown in (b). The red dash-dotted curve represents the lineout of the field $E_x$ in $z=0$ \citep{yao2022detailed}.}
    \label{fig:pr}
\end{figure}

The single shock front was also probed with protons in order to measure the electric field at that location \citep{yao2022detailed}. The probing TNSA protons were sent parallel to the magnetic field, as can be seen in Fig.~\ref{fig:setup} (a). As shown in Fig.~\ref{fig:pr} (a), we could clearly observe the same structure as in Fig.~\ref{fig:setup} (b), namely the front of the expanding plasma and a particular proton deflection structure as the shock front. From it, as shown in Fig.~\ref{fig:pr} (b) and (c), we can infer that the electric field at the shock front had a double jump structure, with a field reversal. Detailed discussions can be found in \citep{yao2022detailed}.

The main parameters of the shock we can produce with our setup are summarized in Table~\ref{tab:solar_para}, showing that they are collisionless and super-critical, and actually quite close to that of the solar wind. Yet, reaching higher velocity shocks, with applicability to supernova remants (SNR), is not possible with presently accessible high-power lasers, but will be in the near future with e.g. Apollon \citep{burdonov2021characterization}.

\begin{table}
\centering
\begin{tabular}{p{0.3\textwidth} p{0.2\textwidth}<{\centering} p{0.15\textwidth}<{\centering} p{0.15\textwidth}<{\centering}}
\hline
\textbf{}                                       & \textcolor{black}{\textbf{Plasma beta}}   & \textcolor{black}{$\boldsymbol{M_{ms}}$} & $\boldsymbol{\lambda_{mfp}/r_{L,i}}$ \\ \hline
\textbf{\makecell{Our Results}}                 & \textcolor{black}{0.2}                   & \textcolor{black}{3.1}                   & \textcolor{black}{12.2}              \\ \hline
\textbf{\makecell{Earth's Bow Shock}}           & \textcolor{black}{$0.4 - 0.8$}             & \textcolor{black}{2.8}                   & $1.2\times10^8$                      \\ \hline
\textbf{\makecell{Non-relativistic SNR}}        & \textcolor{black}{$> 1$}                 & \textcolor{black}{$20 - 100$}               & $10^4$                               \\ \hline
\end{tabular}%
\caption{Comparison between the parameters of the shocks we are able to produce with our platform, and the ones of the solar wind, as well as those from a non-relativistic SNR \citep[see][p. 10]{yao2021laboratory}. The parameters used for the calculations are: $B = 20$ T, $T_e = 100$ eV, $n_e = 3\times 10^{18}$ cm$^{-3}$, $v_{shock} = 1500$ km/s (see details in texts). The plasma beta is defined as $\beta \equiv P_{therm}/P_{mag}$ i.e. the ratio of plasma pressure to magnetic field pressure. The magnetosonic Mach number $M_{ms} = v_{s}/v_{ms}$, is the ratio between the shock velocity and the magnetosonic velocity ($v_{ms} \equiv \sqrt{v_A^2 + c_s^2}$) in the upstream medium, where $v_A$ is the Alfven velocity and $c_s$ is the ion sound velocity. The fourth column of the Table allows one to evaluate that the shock is not affected by collisions. This requires that the ion–ion collision mean free path ($\lambda_{mfp}$) be large compared to the shocked-ion gyroradius, $r_{L,i}$, which sets the scale of the shock transition. 
}
\label{tab:solar_para}
\end{table}

\subsection{Double magnetized shocks collision}

\begin{figure}
    \centering
    \includegraphics[width=\textwidth]{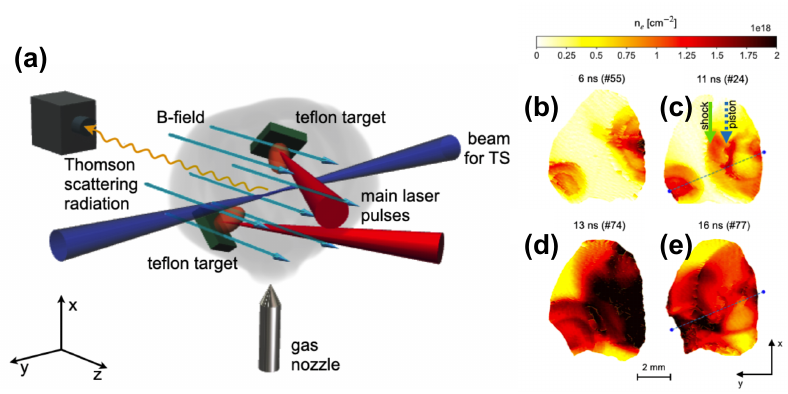}
    \caption{(a) Setup of the experiment with two high-power lasers ((1 ns, 100 J at 1 $\mu$m wavelength, $1.6\times 10^{13}$ W/cm$^2$ intensity on target)) irradiate two solid targets (CF$_2$) embedded in ambient gas and strong B-field (20 T). (b-e) Temporal sequence of the density (along the z-axis) measurements showing the evolution of the collision of two magnetized shocks.}
    \label{fig:double}
\end{figure}

Later, we investigated at LULI2000, the interpenetration of two sub-critical single shocks with a similar setup, which is shown in Fig.~\ref{fig:double} (a). Since the distance between the two targets was large (i.e., 9 mm), due to constraints in the setup, the shocks travel to the middle and collide after 11 ns, as can be seen in Fig.~\ref{fig:double} (c) and (d), thus having an average velocity lower than 500 km/s. This led us to be limited to exploring sub-critical shocks encounter \citep{fazzini2022particle}.

\begin{figure}
    \centering
    \includegraphics[width=\textwidth]{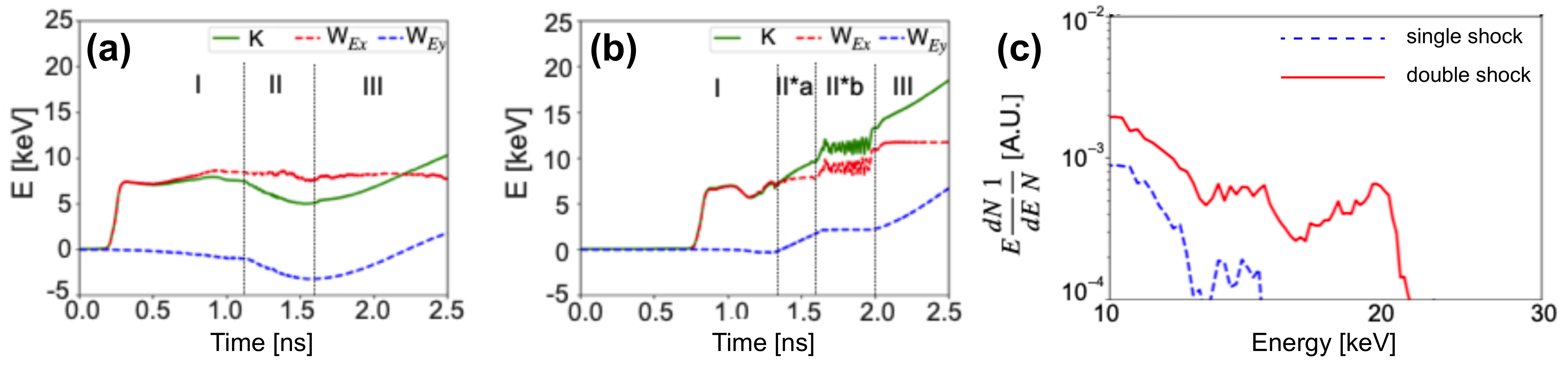}
    \caption{PIC simulation results demonstrating the enhanced particle acceleration in the sub-critical shock colliding case (b), compared to the single one (a). The final energy spectrum is in (c). See details of the three phases in  \cite{fazzini2022particle}. }
    \label{fig:work_slow}
\end{figure}

The measured ion spectrum is the same as shown in Fig.~\ref{fig:setup} (d). Nevertheless, our PIC simulations have revealed an interesting particle dynamics at play during the sub-critical shocks encounter, and which provides energy enhancement (see Fig.~\ref{fig:work_slow}). By tracking a set of representative energetic protons, we have indeed been able to identify that this energy enhancement is a result of the ``phase-locking effect'' between the transverse particle velocity ($V_y$) and the transverse E-field ($E_y$) after the collision, i.e., the work done by $E_y$ (i.e., $W_{Ey}$) is purely positive, which only exists in the shock colliding case and allows a further boost of the ion energy \citep{fazzini2022particle}. 

Unfortunately, the phase-locking effect could not be observed at LULI2000 since the measure time-integrated spectrum of ions is produced by a combination of ions accelerated by the super-critical shock early in time and the collision of the slowed-downed subcritical shock that would produce the phase-locking acceleration. Since, the maximum energy produced by these sub-critical shocks is low, we are not able to separate the two sources in our past measurements. This is why we propose to move, using an improved setup, to investigate the encounter to the moments where the shocks are still super-critical.

\section{Future plans}

As shown above, in our previous work investigating the collision of two sub-critical shocks, we have revealed an interesting effect of phase-locking of the ions trapped between the two counter-streaming shocks, resulting in boosting their energy. 
We have now performed PIC simulations to gauge how this effect would manifest in encountering super-critical shocks. 

\begin{figure}[htp]
    \centering
    \includegraphics[width=\textwidth]{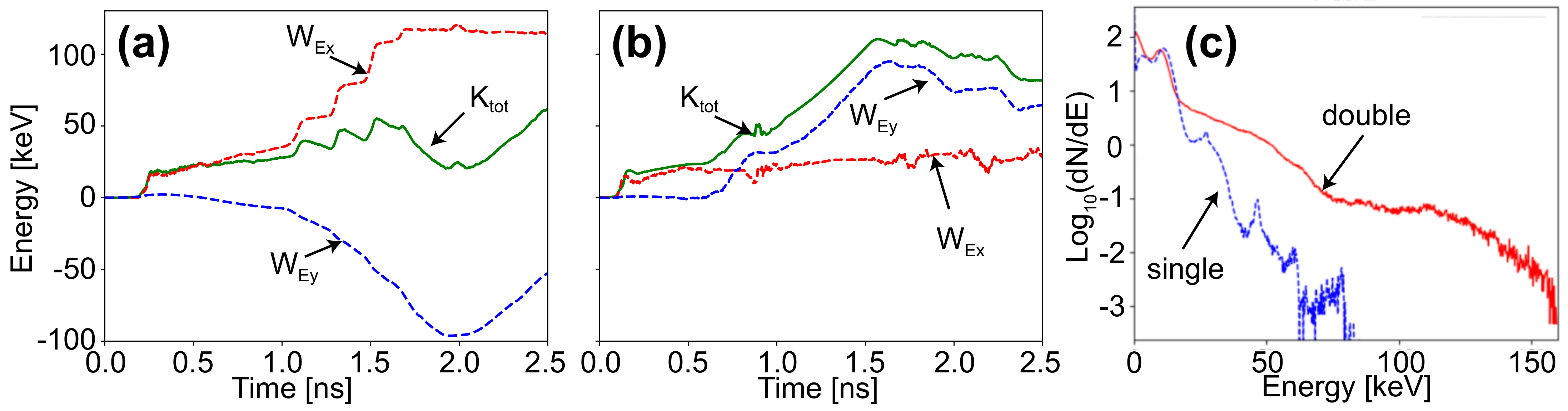}
    \caption{PIC simulation results demonstrating the enhanced particle acceleration in the super-critical shock colliding case (b) we propose to investigate, compared to the single shock one (a), with a shock velocity of 1500 km/s under a 20 T magnetic field. The final energy spectrum is in (c), showing an enhancement of the maximum energy by a factor 2 for the colliding shocks case. Note that now the energy spectrum of the colliding case can be observed in the experiment, since it is higher than the one of the ions accelerated in the single shock, i.e., reaching around 80 keV, see Fig.~\ref{fig:setup} (d).}
    \label{fig:work}
\end{figure}

Fig.~\ref{fig:work} (a) and (b) show the time-evolution of the energy of ions, as well as the contribution from the longitudinal ($W_{Ex}$) and transverse ($W_{Ey}$) electric fields (the energy that they pick up while gyrating in the field) in the case of both a single or of colliding shocks. One can observe that there is, in the colliding case, the same phase-locking effect (i.e., $W_{Ey}$ stays positive due to the shock collision), which induces a significant particle acceleration enhancement. This is shown in the final ion energy spectrum plotted in Fig.~\ref{fig:work} (c). Based on our previous experience in measuring ions energized by single super-critical shocks \citep{yao2021laboratory,yao2022detailed}, this further and strong energization should be easily detectable and recognizable.

\begin{figure}
    \centering
    \includegraphics[width=0.8\textwidth]{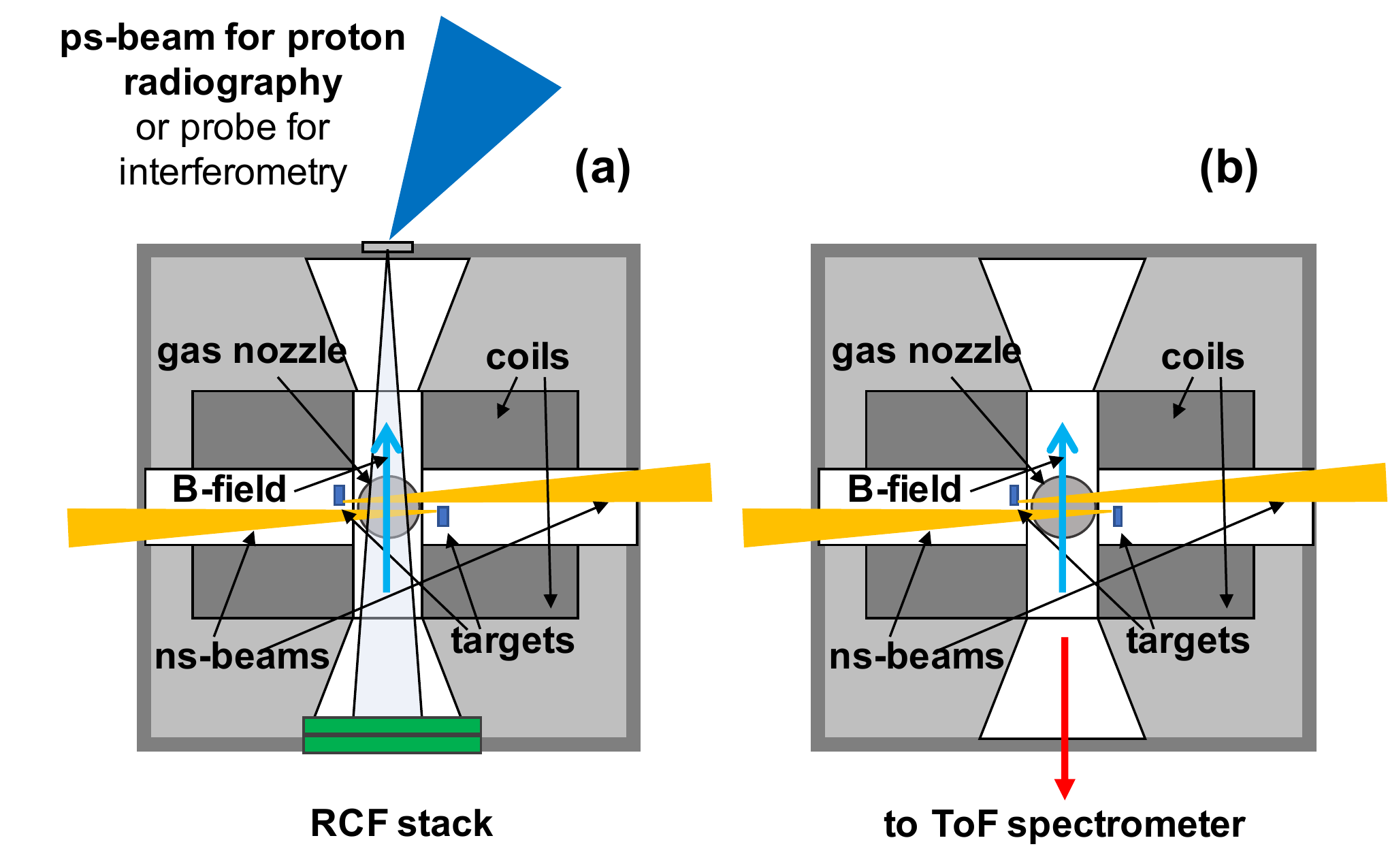}
    \caption{Detailed view inside the coil of the proposed setup. The two ns-laser beams hit the two targets, giving rise to two face-on collisionless shocks, embedded in the 20 T B-field and a $10^{18}$ cm$^{-3}$ ambient gas, that meet at the center of the coil. In setup (a), proton radiography or optical probing will be used along the main B-field axis to diagnose the colliding shocks. Alternatively in setup (b), a ToF spectrometer will be used, also along the B-field axis, to diagnose the dynamics of particle acceleration taking place in the plasma. 
    }
    \label{fig:ral}
\end{figure}

This experiment is programmed to be performed at RAL-TAW in 2023, due to its unique multi-beam capability to generate two super-critical shocks (using two long pulse beams), as well as characterize them in details using optical probing, proton radiography (powered by using the facility high-energy short-pulse B8 beam), space-resolved x-ray spectroscopy, and Thomson scattering. 
We will lift the limitation that was preventing us at LULI2000 (due to a constraint in the laser beams arrangement) to investigate super-critical shock encounter by having closer arrangement of the targets, thus ensuring the shocks meet before 5 ns in their evolution. Fig.~\ref{fig:ral} shows the proposed setup in the horizontal plane. The two targets, from which the two shocks will be launched, will directly face each other, which will also allow for a straight, face-on collision of the shocks. Long f-number long-pulses (in yellow) will irradiate each target and induce the plasma expansion. A pulsed gas jet (coming from the bottom of the coil) will provide the ambient plasma; the gas is ionized by the x-rays from the laser-target interaction. As shown in Fig.~\ref{fig:ral} (a), the short-pulse will be focused on a solid target at the edge of the coil. A TNSA-accelerated proton beam will emerge from the target back side and diagnose the shocks and field structures on a RCF stack. Alternatively to this proton radiography diagnostic, we will set, along the axis of the magnetic field, a time-of-flight (ToF) spectrometer which will measure the time at which ions land on a photomultiplier tube (PMT) \citep{fourmaux2013investigation}, as shown in Fig.~\ref{fig:ral} (b). From it, we will be able to observe the dynamics and the energy of particle acceleration induced within the plasma.

\section{Conclusion}
In summary, by re-using the platform we established at LULI2000 and JLF/TITAN to investigate the dynamics of magnetized shocks and the associated particle acceleration, our future plan will focus on characterizing experimentally the particle acceleration enhancement during the collision of two super-critical shocks. It will serve as the direct proof of our theoretical analysis, possibly confirming the phase-locking effect induced by the shock collision \citep{fazzini2022particle} for the first time in the laboratory. 

\section*{Acknowledgements}

We thank the teams of LULI (France) and JLF (USA) laser facilities for their expert support, as well as the Dresden High Magnetic Field Laboratory at Helmholtz-Zentrum-Dresden-Rossendorf for the development of the pulsed power generator. We thank the SMILEI development team for technical support. This work was supported by funding from the European Research Council (ERC) under the European Unions Horizon 2020 research and innovation program (Grant Agreement No. 787539). 
S.O. and M.M. acknowledge financial contribution from the PRIN INAF 2019 grant ``From massive stars to supernovae and supernova remnants: driving mass, energy and cosmic rays in our Galaxy'' and the INAF mainstream program ``Understanding particle acceleration in galactic sources in the CTA era''.
The computational resources of this work were supported by the National Sciences and Engineering Research Council of Canada (NSERC) and Compute Canada (Job: pve-323-ac, PA).  Part of the experimental system is covered by a patent (1000183285, 2013, INPI-France). 
The research leading to these results is supported by Extreme Light Infrastructure Nuclear Physics (ELI-NP) Phase II, a project co-financed by the Romanian Government and European Union through the European Regional Development Fund, and by the project ELI-RO-2020-23 funded by IFA (Romania).

\bibliographystyle{jpp}
\bibliography{refs}

\end{document}